\documentclass[onecolumn,groupedaddress,amsmath,preprint,showpacs,aps,pra]{revtex4-1}

\usepackage{graphicx}

\begin{document}
\title{Measurement of the Kerr nonlinear refractive index of Cs vapor}

\author{Michelle O. Ara\'{u}jo, Hugo L. D. de S. Cavalcante, Marcos Ori\'{a}, Martine Chevrollier and Thierry Passerat de Silans.}
\email{thierry@otica.ufpb.br}
\affiliation{Laborat\'{o}rio de Espectroscopia \'{O}tica, DF-CCEN, Cx. Postal 5086 - Universidade Federal da Para\'{i}ba, 58051 - 900 Jo\~{a}o Pessoa - PB, BRAZIL}
\author{Romeu Castro, Danieverton Moretti.}
\affiliation{UAF, Universidade Federal de Campina Grande, 58051-970 Campina
Grande, PB, Brazil}
\date{\today}

\begin{abstract}
Atomic vapors are systems well suited for nonlinear optics studies but very few direct measurements of their nonlinear refractive index have been reported. Here we use the z-scan technique to measure the Kerr coefficient, $n_2$, for a Cs vapor. Our results are analyzed through a four-level model, and we show that coherence between excited levels as well as cross-population effects contribute to the Kerr-nonlinearity.
\end{abstract}
\maketitle

\section{INTRODUCTION}

Atomic vapors are systems well suited for studies of optical nonlinearities. First of all, they are easy to saturate, which enables the observation of nonlinear effects with low intensity continuous-wave laser light \cite{Ashkin74,Suter93}. At the same time, atomic vapors are damage-free which is important, for instance, for filamentation studies \cite{Abraham90}. Second, as the resonances are sharp the nonlinear parameters can be easily modified by finely tuning the frequency near or across a resonance \cite{Boller91}. This allows to play with the relative contributions of linear and nonlinear effects by changing the laser wavelength. Third, atomic systems allow for a variety of level schemes exploring fine, hyperfine and Zeeman levels such as: two-level systems \cite{Labeyrie2003,Saffman04}, $\Lambda$ three-level schemes \cite{Boller91}, double-$\Lambda$ four-level schemes \cite{Lukin00,Passerat11}, five-level schemes \cite{Andersen01} and so on. Fourth, in most experiments, when one can ignore radiation trapping and collisional effects, atomic vapors behave as locally saturable media and are thus easy to model \cite{Ackemann98}.\\

As atomic vapors are isotropic media, the first nonlinear contribution to the polarization is a third-order term in the electric field ($\chi^{(3)}E^3$), in the dipole approximation \cite{Chi2}. The third-order susceptibility $\chi^{(3)}$ is responsible for phenomena such as EIT \cite{Boller91}, four-wave mixing \cite{Liao78}, third-harmonic generation \cite{Harris71}, self-focusing and self-trapping effects \cite{Chiao64,Grischkowsky70}. Those phenomena are expected to have applications, for instance, in correlated photons generation \cite{Lett09}, nondemolition measurement \cite{Xiao08} and generation of optical solitons \cite{Suter93}. In this article, we are interested in the self-focusing of a light beam, which originates from the real part of the third-order susceptibility and results in a Kerr-like term in the medium refraction index: $n=n_0+n_2I$. Self-focusing was observed in the early seventies \cite{Grischkowsky70}. The change from self-focusing to self-defocusing behavior when one scans the laser frequency through an atomic resonance has recently been used to generate an error signal for frequency stabilization \cite{ANGELLS,ANGELLS2}.\\

A very simple and easy-to-implement technique to measure the Kerr coefficient, $n_2$, is the well known z-scan technique \cite{Sheik}. Despite the potential of atomic samples for self-focusing study, very few direct experimental measurements of $n_2$ have been carried on. The z-scan technique was used to probe Na \cite{Sinha02} and Rb \cite{Mccormick04,Mccormick03} vapors and for clouds of cold Cs atoms \cite{Saffman04}. To our knowledge, no measurements have been made exploring the D2 line of hot Cs vapors, on which we report in this article. As alkali atomic vapors have sharp resonances, with linewidths of a few MHz for cold atomic clouds to hundreds of MHz for Doppler-broadened resonances, the behavior of $n_2$ with detuning is rich and accessible to lasers with relatively narrow tuning ranges. This makes atomic systems qualitatively different from solid-state and liquid systems, these two last exhibiting nonlinear properties varying weakly with frequency. In \cite{Sinha02} are reported measurements in a sample of Na vapor, carried out far from central resonance. In this limit, $n_2$ has a well-known behavior $n_2\sim1/\delta^3$ \cite{Grischkowsky70}, where $\delta$ is the frequency detuning. Experimental results for detunings of a few Doppler widths from Rb resonance \cite{Mccormick04} indicate that a model of velocity-integration of the resonant lineshape, simulated by the derivative of a Gaussian function, is more adequate than the $1/\delta^3$ behavior. There are no reports on the observation of the two regimes of detuning in the same system. Moreover, the expressions used in \cite{Sinha02,Mccormick04} are derived from a two-level model, which is a reasonable approximation in these systems where the Doppler width is much larger than the hyperfine spacing. However, as the Cs 6P$_{3/2}$ hyperfine sublevel spacing is close to the Doppler width of the D2 transition, one has to consider a four-level system in order to get a more realistic description. In this article we measure $n_2$ for a hot Cs vapor in both the close-to- and the far-from-resonance regimes and develop a four-level model, consisting in one ground and three excited levels. We show that cross-population and excited coherence terms give important contributions to the $n_2$ value. The experimentally measured $n_2$ values are shown to be much more consistent with this four-level theory.

\section{THEORETICAL MODELS}
Self-focusing of a laser beam in a nonlinear medium is a third-order nonlinear effect, i.e., it is induced by the intensity-dependent term in the refractive index $n=n_0+n_2I$. The Kerr coefficient, $n_2$, is related to the third-order susceptibility, $\chi^{(3)}$ by \cite{Boyd}:
\begin{equation}
n_2=\frac{3}{4n^2_0\epsilon_0c}\Re{\chi^{(3)}}, \label{n2Chi}
\end{equation}
where $n_0$ is the intensity-independent refractive index ($n_0\approx1$ for a dilute vapor), $\epsilon_0$ is the vacuum permittivity, $c$ is the speed of light in vacuum and $\Re{\chi^{(3)}}$ is the real part of $\chi^{(3)}$.\\

The third-order susceptibility can be calculated, using the density matrix formalism, as:
\begin{equation}
\chi^{(3)}=\frac{N}{3E^3\epsilon_0}\sum_j\left(\mu_{j0}\rho^{(3)}_{0j}+\mu_{0j}\rho^{(3)}_{j0}\right), \label{chi3}
\end{equation}
where $\left|j\right\rangle$ denotes the excited states, $\left|0\right\rangle$ denotes the ground state, $N$ is the atomic density, $E$ is the electric field amplitude and $\mu_{0j}=\left\langle 0\right|\hat{\mu}\left|j\right\rangle$ is the ground-excited electric dipole matrix element. In (\ref{chi3}), the density matrix has been expanded in a series of powers of $E$: $\rho=\sum\limits_{l}\rho^{(l)}$ and $\rho^{(3)}_{0j}$ is the ground-excited coherence term that has a cubic dependence with $E$.\\

For an atomic vapor, one has to integrate the velocity-dependent coherence term, $\rho_{0j}(v)$, over the Maxwell-Boltzmann velocity distribution, $W(v)$, to take into account the Doppler shift induced by the atomic motion. Thus,
\begin{equation}
\rho_{0j}=\int dv\: W(v)\rho_{0j}(v).
\end{equation}

In the following, we will calculate the Kerr coefficient first for a general two-level system, and then for the specific Cs D2 line, for which we take into account one hyperfine ground state and three hyperfine excited levels.\\

\subsection{Two-level system}

We write a Hamiltonian for a closed two-level system in the rotating-wave and dipole approximations, which is given by:
\begin{equation}
H=\hbar\omega_j\left|j\right\rangle\left\langle j\right|-\hbar\Omega_j e^{i\omega t}\left|0\right\rangle\left\langle j\right|-\hbar\Omega_j e^{-i\omega t}\left|j\right\rangle\left\langle 0\right|, \label{twoH}
\end{equation}

where $\hbar\omega_j$ is the energy of the excited state $\left|j\right\rangle$ (the ground state is taken at zero energy), $\Omega_j=\mu_{0j} E/\hbar$ is the Rabi frequency and $\omega$ is the field frequency.

For an atom with velocity component $v$ along the beam axis, it is well known that the real part of the susceptibility can be written as \cite{Boyd}:
\begin{equation}
\Re{\chi_v}=\frac{4N\mu_{0j}}{E	\epsilon_0}\frac{\Omega_j\delta_v/\Gamma^2}{\left(1+\frac{4\delta_v^2}{\Gamma^2}+\frac{8\Omega^2}{\Gamma^2}\right)^2}, \label{2levelChi}
\end{equation}
where $\delta_v=\omega-\omega_j-kv=\delta-kv$, $k$ is the wavenumber and $\Gamma$ is the homogeneous linewidth. For a weak light intensity one can expand the expression (\ref{2levelChi}) and obtain \cite{Wang04}:
\begin{equation}
\Re{\chi^{(3)}_v}=\frac{32N\mu_{0j}^4}{3\epsilon_0\hbar^3}\frac{\delta_v/\Gamma^4}{\left(1+\frac{4\delta_v^2}{\Gamma^2}\right)^2}, \label{2levelRChi}
\end{equation}
which gives the contribution to the third-order susceptibility from atoms in each class of velocity.  To sum the contributions of all the atoms of the vapor, one integrates (\ref{2levelRChi}) over the Maxwell-Boltzmann distribution of atomic velocities:
\begin{equation}
\Re{\chi^{(3)}}=\int dv \Re{\chi^{(3)}_v}W(v). \label{2levelIntegration}
\end{equation}

Notice that $\chi^{(3)}$ has a strong spectral dependence around the frequency of atomic transitions. Therefore, we will now consider two asymptotic regimes for the velocity integration: i) close to resonance, and under the condition $\Gamma\ll\Gamma_D$ ($\Gamma_D$ is the Doppler width), and ii) far from resonance.\\

Close to resonance, the main contribution to the integral (\ref{2levelIntegration}) comes from the classes of velocity for which the detuning is given by $\delta_v=\pm\Gamma/\sqrt{12}$ in the atomic reference frame (maximum of expression (\ref{2levelRChi})). Thus, $n_2$ is proportional to the difference of population densities for which $\delta-kv=\pm\Gamma/\sqrt{12}$. Therefore, $n_2$ is proportional to the derivative of a Gaussian lineshape \cite{Mccormick04}:
\begin{equation}
n_2(\text{cm}^2\text{/W})=10^4\times\frac{8\pi^{7/2}\mu_{0j}^4N}{3c\epsilon_0^2h^3}\frac{2\pi\delta}{\Gamma(ku)^3} e^{-4\pi^2\delta^2/(ku)^2},\label{Close}
\end{equation}
where $u$ is the width of the atomic velocity distribution.\\

For a radiation tuned far from atomic resonance, all velocity classes that have a sizable population comply with the condition $\left|\delta_v\right|\gg0$. Thus, the contribution of all the atoms is essentially nonresonant, and all the velocity classes contribute in the same way (weighted by the population density) to the integral (\ref{2levelIntegration}). The Kerr coefficient is then given by the far-from-resonance limit of expression (\ref{2levelRChi}) and exhibits the well known $\delta^{-3}$ behavior ($\delta_v\gg\Gamma_D$) \cite{Sinha02,Mccormick04}:
\begin{equation}
n_2(\text{cm}^2\text{/W})=10^4\times\frac{\mu_{0j}^4N}{2c\epsilon^2_0\hbar^3\delta^{3}}. \label{far}
\end{equation}

\begin{figure}
	\centering
		\includegraphics{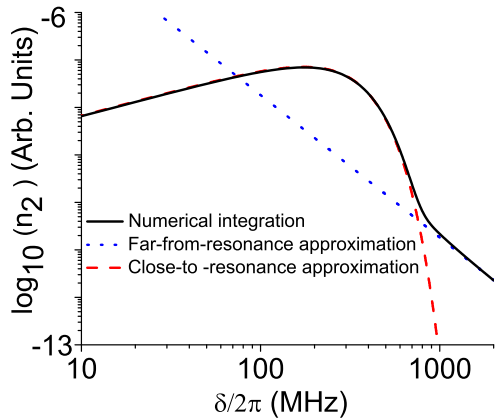}
	\caption{(Color online) Calculated $n_2$ values (two-level model, Eq.(\ref{2levelIntegration})) as a function of the laser detuning and the close- (red dashed, Eq. (\ref{Close})) and far-from-resonance (blue dots, Eq. (\ref{far})) asymptotic behavior, in a log$_{10}$-log$_{10}$ scale.}
	\label{fig:DoisniveisIntegracao}
\end{figure}

To obtain $n_2$ in the intermediate detuning range, one has to integrate equation (\ref{2levelIntegration}). In order to compare the obtained result with the two asymptotic expressions (\ref{Close}) and (\ref{far}), we show in figure \ref{fig:DoisniveisIntegracao} the $n_2(\delta)$ curve numerically calculated from (\ref{2levelIntegration}), for a large detuning range on the blue side of the resonance, together with the close (Eq. (\ref{Close})) and far (Eq. (\ref{far})) from resonance asymptotic curves. Similar results are obtained on the red side of the resonance since $\chi^{(3)}$ has an antisymmetric lineshape with detuning in a two-level model. To our knowledge, previous measurements of $n_2$ have been limited to one of the asymptotic regimes \cite{Sinha02,Mccormick04}, and no one has explored all the detuning range. The observation of the Kerr coefficient in a large range of detunings is one of the accomplishments reported in this article.

\subsection{Multi-level system}

\begin{figure}[htbp]
	\centering
		\includegraphics{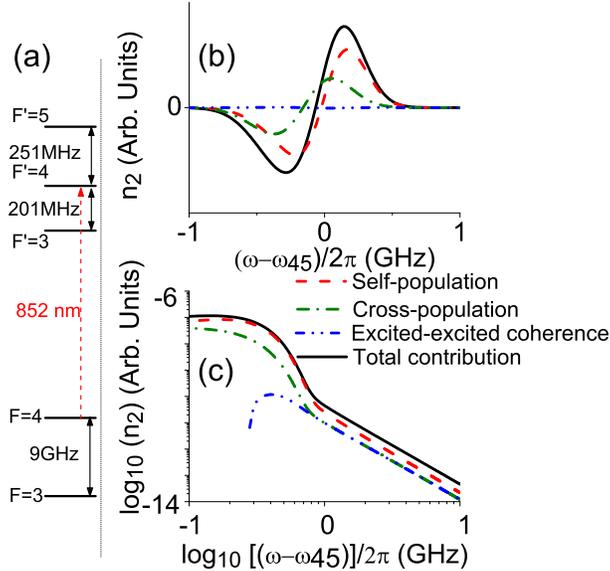}
	\caption{(Color online) a) Schematic representation of the relevant sub-levels of the Cs $6S_{1/2} (F=4)\rightarrow 6P_{3/2}(F'=3,4,5)$ transition (out of scale). b) Calculated values of $n_2$ using the ground-excited coherence from the four-level model (equation \ref{rho3}). The contributions of the self-population (dashed, red), cross-population (dots, green) and the coherence between excited levels (dot-dashed, blue) are shown, together with the total Kerr coefficient (solid, black). c) Same as b) for blue-detuned frequencies relative to the Cs $6S_{1/2}(F=4)\rightarrow 6P_{3/2}(F'=5)$ transition, in log$_{10}$-log$_{10}$ scale.}
	\label{fig:Fig2}
\end{figure}

The D2 transition of alkali atoms has multiple excited hyperfine levels. In Cs atoms, the hyperfine energy splitting is of the same order of magnitude as the Doppler width. As a consequence, this hyperfine structure must be taken into account in the $n_2$ lineshape calculation. On the other hand, the splitting between fundamental hyperfine levels is much larger than the typical laser linewidth and the Doppler width of the D2 transition, so that we only take one ground state into account. Therefore, we consider cesium atoms as closed four-level systems consisting of one fundamental hyperfine level and three dipole-accessible excited hyperfine levels of the D2 transition (see Figure \ref{fig:Fig2}a). The Hamiltonian considered here is a generalization of the Hamiltonian written above (eq. \ref{twoH}) for the two-level system:
\begin{equation}
H=\hbar\sum_j\omega_j\left|j\right\rangle\left\langle j\right|-\sum_j\left[\hbar\Omega_j e^{i\omega t}\left|0\right\rangle\left\langle j\right|-\hbar\Omega_j e^{-i\omega t}\left|j\right\rangle\left\langle 0\right|\right],
\end{equation}

and its expansion in powers of $I$ is not straightforward. To gain a direct insight into $\chi^{(3)}$, we consider a perturbative expansion of the density matrix elements $\rho=\sum\limits_{l}\rho^{(l)}$ (see Appendix), and calculate $\chi^{(3)}$ using (\ref{chi3}). The steady-state solutions for this third-order density matrix ground-excited coherence are given by:
\begin{equation}
\rho^{(3)}_{0j}=\frac{-i\delta_j-\Gamma/2}{\delta_j^2+\Gamma^2/4}\left[2i\Omega_j\rho^{(2)}_{jj}+i\sum_{l\neq j}\Omega_j\rho^{(2)}_{ll}+i\sum_{l\neq j}\Omega_l\rho^{(2)}_{lj}\right] \label{rho3}
\end{equation}

Analyzing the contribution to $n_2$ (eq. \ref{n2Chi}) of the first term inside the brackets in equation (\ref{rho3}), we notice that it simply consists in the summation of three independent two-level systems. Since the electric dipole moments are different for every hyperfine transition, the resulting lineshape is slightly asymmetric, as depicted in Figure \ref{fig:Fig2}. The $F=4\rightarrow F'=5$ contribution dominates because of its larger dipole moment, since $n_2$ scales as $\mu^4$. This first term is the index effect of population exchange between the ground state and the excited level $\left|j\right\rangle$ and we call it the {\it self-population} contribution. The change in population in the other excited states is at the origin of the second term inside the brackets in equation (\ref{rho3}) and we call it the {\it cross-population} contribution. This term results from the ground state depopulation and enhances the $n_2$ values, modifying the lineshape towards a more symmetric shape than the self-population term. The third term inside the brackets in equation (\ref{rho3}) comes from a coherence build-up between excited states \cite{Asadpour12}, and its relative contribution to $n_2$ is greater at large detunings (see Figure \ref{fig:Fig2}c).\\

In the far-from-resonance asymptotic regime all terms in equation (\ref{rho3}) scale as $\delta^{-3}$ and one obtains back the same $n_2$ values given by expression (\ref{far}). In this limit, the self-population contributes one-half of the signal while the cross-population and the excited coherence terms contribute one-fourth each.\\

\section{EXPERIMENT}
\begin{figure}
	\centering
		\includegraphics{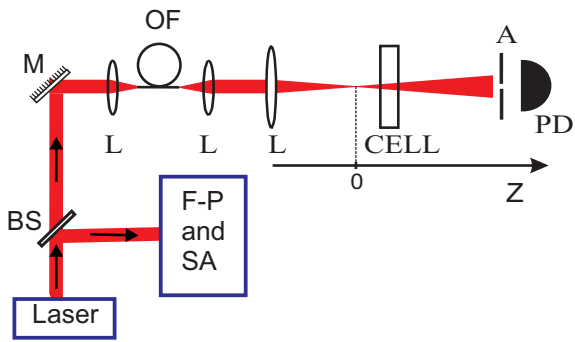}
	\caption{(Color online) Experimental setup. The laser beam passes through a single-mode optical fiber to generate a fundamental Gaussian profile. After the fiber, the beam passes through a 20-cm-focal lens and is detected in the far field region after a circular aperture, placed 39 cm from the focus point. A Cs vapor cell of thickness 1 mm is scanned along the beam path ($z$-axis). A Fabry-Perot interferometer (F-P) and a saturated absorption (SA) set-up allow to monitor the laser frequency. OF is an optical fiber, L are lenses, M is a mirror, BS is a beamspliter and PD is a photodetector.}
	\label{fig:Fi3}
\end{figure}

We measured the Kerr nonlinearity of a hot Cs vapor with a setup of the well known z-scan technique \cite{Sheik}. The experimental configuration is shown in Fig. (\ref{fig:Fi3}). We collimate a Gaussian beam of diameter $3$ mm at the output of a monomode fiber. This beam is then focused by a 20-cm-focal lens and detected in the far field after the focal point. The shape of the beam and its diameter are checked all along the beam path using the knife-edge technique \cite{Arnaud71,Khosrofian83}. The beam is linearly polarized. No magnetic shielding is used, nor is applied any external magnetic field, so that the system is submitted to the geomagnetic field only. An aperture is placed before the detector to spatially filter the beam. The far-from-resonance aperture transmittance is $S=0.6$. When a 1-mm-thick cell containing Cs vapor is displaced along the beam accross the focal point, the light intensity transmitted through the aperture is modified due to self-focusing/defocusing effects in the vapor. The cell thickness is shorter than the Rayleigh length ($\sim 5$ mm) so that beam shaping due to propagation and nonlinear refraction in the vapor is negligible \cite{Sheik}. The aperture transmittance as a function of the cell position $z$, relative to the focal point, is given by \cite{Sheik}:\\
\begin{equation}
T=1-\frac{4\,\Delta\Phi_0 x}{\left(x^2+9\right)\left(x^2+1\right)},\label{transmission}
\end{equation}
where $x=z/z_R$, $z_R$ is the Rayleigh length and $\Delta\Phi_0$ is the on-axis phase shift at focal point \cite{Sheik, Mccormick04}.\\

Fitting equation (\ref{transmission}) to the experimental curve one obtains the Kerr coefficient through:
\begin{equation}
n_2=\frac{\lambda}{2\pi I_0L}\Delta \Phi_0
\end{equation}
where $\lambda$ is the light beam wavelength, $L$ is the cell thickness and $I_0$ is the light intensity at focal point. The on-axis phase shift is proportional to the peak-to-peak amplitude of the transmission signal ($\Delta T$) \cite{Sheik}, 
\begin{equation}
\Delta\Phi_0=\Delta T/\left[0.406\left(1-S\right)^{0.25}\right] \label{DT}.
\end{equation}

\section{EXPERIMENTAL RESULTS}
For the slightly heated Cs vapor we used ($T\approx70^\circ C$ , $N=2.4\times10^{12}$ atoms/cm$^3$ \cite{density}), the Doppler width is $\Gamma_D\sim2\pi\times250\times10^6\:\text{s}^{-1}$ . We have measured $n_2$ for red detuning $\left|\delta\right|\geq 2\pi\times600\times10^6\:\text{s}
^{-1}$ = 2.4 $\Gamma_D$ relative to the cyclic hyperfine transition. The intensity transmitted through the aperture is affected by defects of the moving elements. Thus, we have normalized the z-scan signal at the frequencies of interest by the $z$-scan signal at a frequency detuned $4$ GHz to the red side of the resonance. Furthermore, for the range of detuning from $600$ MHz to $800$ MHz, the nonlinear absorption is not negligible and we have further normalized the signal of the aperture transmission by an open-aperture signal. This procedure showed to be enough to obtain good values of $n_2$, even though, for a rigorous approach one should take into account the attenuation of the intensity through the vapor for those frequencies (the linear absorption ranges from $40\%$ for $\delta=2\pi\times600\times10^6\:\text{s}^{-1}$ to $10\%$ for $\delta=2\pi\times800\times10^6\:\text{s}^{-1}$) \cite{Mccormick03}.\\

\begin{figure}
	\centering
		\includegraphics{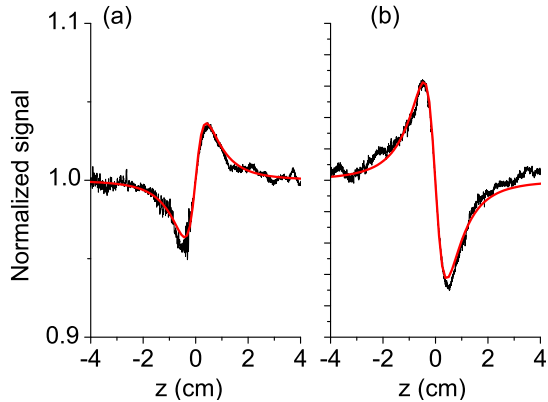}
	\caption{(Color online) Z-scan curve for (a) red detuning, $\omega-\omega_{45}=-2\pi\times1000\times10^6\:\text{s}^{-1}$ and $N=2.8\times10^{12}$ atoms/cm$^3$; (b) blue detuning, $\omega-\omega_{45}=2\pi\times600\times10^6\:\text{s}^{-1}$ and $N=2.4\times10^{12}$ atoms/cm$^3$. Black curves: experimental data. The red curves are best fits to the experimental curves and are calculated from equation (\ref{transmission}).}
	\label{fig:Fig4}
\end{figure}

In figure \ref{fig:Fig4} we show two typical normalized z-scan curves obtained when the laser is detuned to the red side (Fig. \ref{fig:Fig4}a) or to the blue side (Fig. \ref{fig:Fig4}b) of the resonance.  Note that, typically, a change of a few percent in the aperture transmission is obtained. For red detunings ($n_2<0$), the medium is self-defocusing and, as a consequence, the aperture transmission is increased when the cell is before the focal point and diminished when the cell is beyond it (Fig. \ref{fig:Fig4}a). Conversely, for blue-detuned laser frequencies ($n_2>0$), the medium is self-focusing and a decrease followed by an increase of the signal is observed when the cell goes through the laser focus (Fig. \ref{fig:Fig4}b). The signals were fitted using equation (\ref{transmission}) and the fit parameters allow to obtain values of $n_2$.\\

For equation (\ref{transmission}) the condition $\Delta\Phi_0\ll1$ must be fulfilled while for equation (\ref{DT}) $\Delta\Phi_0<\pi$ gives good enough values \cite{Sheik}. In our experiment the maximum value of $\Delta\phi_0$ is $0.4$ which gives good measured $n_2$ values. Ideally, one should use low-intensity beams to avoid higher order effects in the refractive index expansion ($n_4I^2$). Nevertheless, as $n_2$ decreases rapidly with detuning, the signal-to-noise ratio becomes small for low-intensity beams ($\Delta\Phi_0\propto n_2I\rightarrow0$). We have thus repeated the measurements for a few intensity values and fitted the measured $n_2(I)$ by a saturation law $n_2(I)=n_{2}^{ns}/(1+I/I_S)$, where $n_2^{ns}$ is the desired non-saturated Kerr coefficient value and $I_S$ is the detuning-dependent saturation intensity, which is kept as a fit parameter.\\

\begin{figure}
	\centering
		\includegraphics{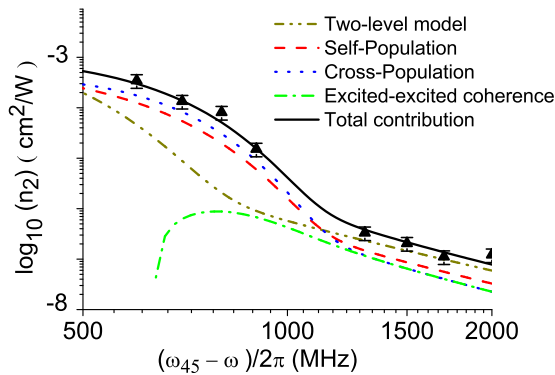}
	\caption{(Color online) Values of $n_2$ for red detuned frequencies relative to $F=4\rightarrow F'=5$ transition, and $T=70^\circ C$ ($N=2.4\times10^{12}$ atoms/cm$^3$). Triangle: experimental data. Other curves: different contributions to the theoretical calculation (see equations (\ref{chi3}) and (\ref{rho3})) together with the Kerr coefficient calculated for a two level model $F=4\rightarrow F'=5$. The error bars are estimated to be $30\%$ of $n_2^{ns}$ values.}
	\label{fig:ResultsExp}
\end{figure}

We have repeated the z-scan measurements for a variety of detunings and plotted the respective values of $n_2^{ns}$ as a function of $\delta$  in Figure \ref{fig:ResultsExp}. Two asymptotic regimes are clearly identified: for small detunings, $n_2^{ns}$ follows the derivative of a Gaussian lineshape, while, for large detunings, a $\delta^{-3}$ dependence is observed. In Figure \ref{fig:ResultsExp}, we also show the theoretical values of $n_2$, calculated from the ground-excited coherence given by equation (\ref{rho3}), as well as the particular contribution of each term separately. The theoretical $n_2$ curve fits well the experimental values, showing that considering simply a summation of independent two-level models is not enough to accurately calculate $n_2$.\\ 

We have estimated an upper limit for the error, of 30\% of the nonsaturated $n_2$ values. Splitting between Zeeman sublevels can be ignored since the maximum magnetic splitting due to geomagnetic field is smaller than 1 MHz. However, the Zeeman structure modifies the atom-field interaction, introducing, for instance, optical pumping between sublevels, which results in changes in the saturation intensity \cite{Mccormick03}. The good agreement between experiment and calculated values indicates that the contribution of the Zeeman structures to the signal is inside the errors bars. The difference between the values of $n_2$ obtained with the two-level and the four-level models is larger than the error bars for detunings below 1600 MHz.  

The ratio between measured $n_2^{ns}$ values and the vapor atomic densities ranges from $n_2/N=1.5\times10^{-16}$ cm$^5$/W for detunings of the order of two Doppler widths to $n_2/N=5\times10^{-20}$ cm$^5$/W for large detunings. The obtained values are comparable to the ones obtained in \cite{Mccormick04} for a Rb vapor, $n_2/N=10^{-19}$ cm$^5$/W for a detuning of 1 GHz. 

\section{CONCLUSION}
We have measured the Kerr coefficient for a Cs vapor for a large range of frequencies. The obtained $n_2^{ns}$ values vary over four decades as a function of the laser detuning. The experimental results clearly show two asymptotic regimes: a lineshape as the derivative of a Gaussian-like curve for detunings of the order of two times the Doppler widths, and a $\delta^{-3}$ behavior for much larger detunings. To interpret these asymptotic behaviors, the velocity integration for a two-level model was used and showed that it is not accurate for the prediction of $n_2$ values on the full detuning range. We have used a four-level model (one ground and three excited hyperfine levels) that correctly predicts the experimental results. From this multilevel model we showed that cross-population contribution and the build up of coherence between excited levels must be taken into account to accurately calculate $n_2$. Further refinement of the theory, such as considering the Zeeman structure, does not seem to be necessary for the level of measurement precision we have.

\begin{acknowledgments}
This work was partially funded by Conselho Nacional de Desenvolvimento Cient\'{i}fico e Tecnol\'{o}gico (CNPq, contract 472353/2009-8, 470834/2012-9 and 484774/2011-5), Coordena\c{c}\~{a}o de Aperfei\c{c}oamento de Pessoal de N\'{i}vel Superior (CAPES/Pr\'{o}-equipamentos) and Financiadora de Estudos e Projetos (FINEP).
\end{acknowledgments}

\section*{APPENDIX: DETAILS OF THE MULTI-LEVEL CALCULATIONS}

The Hamiltonian of the system is written in the dipole and in the rotating wave approximation as:
\begin{equation}
H=\hbar\sum_j\omega_j\left|j\right\rangle\left\langle j\right|-\sum_j\left[\hbar\Omega_j e^{i\omega t}\left|0\right\rangle\left\langle j\right|-\hbar\Omega_j e^{-i\omega t}\left|j\right\rangle\left\langle 0\right|\right],
\end{equation}
where the Rabi frequencies are written $\Omega_j=\mu_{0j}E/\hbar$. The Zeeman sublevels are not taken into account in our model and the matrix elements of the electric dipole moment are taken between the ground hyperfine level $F=4$ and the excited hyperfine levels $F'=3, 4, 5$.\\

The matrix elements of the electric dipole moment are calculated as \cite{Steck}:
\begin{equation}
\mu_{0F'}=\frac{1}{3}\left(2F'+1\right)\left(2J+1\right)\left\{\begin{array}{ccc}
J & J' & 1\\
F & F' & I
\end{array}\right\}^2\left|\left\langle J\right|\mu\left|J'\right\rangle\right|^2,
\end{equation}
where $F(J)$ and $F'(J')$ represent the total atomic (electronic) angular momentum quantum numbers for ground and excited levels, respectively, and the term inside the brackets is the Wigner 6-j symbol.\\

The fine-structure electric dipole moment is:
\begin{equation} 
\left|\left\langle J'\right|\mu\left| J'\right\rangle\right|^2=\frac{3\pi\epsilon_0\hbar c^3}{\omega_0^3\tau}\frac{2J'+1}{2J+1},
\end{equation}
where $\tau$ is the excited state lifetime. \\

In order to obtain the third-order atomic susceptibility we calculate the density matrix using perturbation theory. This is done by substituting $\rho$ into the density matrix equation of motion by $\sum_{k=0}\lambda^N\rho^N$ and $V$ by $\lambda V$. $\lambda$ is a parameter with values between zero and one and $V=\sum_j\left[\hbar\Omega_j e^{i\omega t}\left|0\right\rangle\left\langle j\right|-\hbar\Omega_j e^{-i\omega t}\left|j\right\rangle\left\langle 0\right|\right]$ is the interaction potential, treated as a perturbation. Equating the terms with the same power of $\lambda$ one obtains:\\

\begin{eqnarray}
\rho_{mn}^{(0)}=-\frac{i}{\hbar}\left[H_0,\rho_{mn}^{(0)}\right]+\textrm{r.t.}\\
\rho_{mn}^{(k)}=-\frac{i}{\hbar}\left[H_0,\rho_{mn}^{(k)}\right]-\frac{i}{\hbar}\left[V,\rho_{nm}^{(k-1)}\right]+\textrm{r.t.},
\end{eqnarray}

where $H_0=H-V$ and r.t. are relaxation terms. 
 
In the zero-order density matrix (without light field), the only non-zero term is $\rho^{(0)}_{00}=1$. For the first-order density matrix one obtains the usual linear result:
\begin{equation}
\rho^{(1)}_{0j}=\frac{i\Omega_j}{i\left(\omega_j-\omega\right)-\Gamma/2}
\end{equation} 
Thus, in the linear regime, the four-level system is equivalent to the sum of three independent two-level systems \cite{Siddons08}.\\

For the second-order density matrix the ground-excited coherence term is zero as it is expected for isotropic media, while the population terms and the excited-excited coherences are non-zero. The population and excited-excited coherences, that only appear in the non-linear regime, are responsible for the second and third terms in the right side of equation (\ref{rho3}).


\begin{thebibliography}{} 
\bibitem{Ashkin74} J. E. Bkorkholm and A. Ashkin, Phys. Rev. Lett. \textbf{32}, 129 (1974).
\bibitem{Suter93} D. Suter and T. Blasberg, Phys. Rev. A \textbf{48}, 4583 (1993).
\bibitem{Abraham90} N. B. Abraham and W. J. Firth, J. Opt. Soc. Am. B \textbf{7}, 951 (1990).
\bibitem{Boller91} K.-J. Boller, A Imamo$\breve{g}$lu, and S. Harris, Phys. Rev. Lett. \textbf{66}, 2593 (1991).
\bibitem{Labeyrie2003}G. Labeyrie, T. Ackemann, B. Klappauf, M. Pesch, G.L. Lippi, and R. Kaiser, Eur. Phys. J. D \textbf{22}, 473 (2003).
\bibitem{Saffman04} Y. Wang and M. Saffman, Phys. Rev. A \textbf{70}, 013801 (2004).
\bibitem{Lukin00} M. D. Lukin, P. R. Hemmer, and M. O. Scully, Adv. At. Mol. Opt. Phys. \textbf{42}, 347 (2000).
\bibitem{Passerat11} T. Passerat de Silans, C.S.L. Gon\c calves, D. Felinto, and J. W. R. Tabosa, J. Opt. Soc. Am. B \textbf{28} 2220 (2011).
\bibitem{Andersen01} J.A. Andersen, M. E. J. Friese, A. G. Truscott, Z. Ficek, P. D. Drummond, N. R. Keckenberg, and H. Rubinztein-Dunlop, Phys. Rev. A \textbf{63}, 023820 (2001).
\bibitem{Ackemann98} T. Ackemann, T. Scholz, Ch. Vorgerd, J. Nalik, L. M. Hoffer, and G. L. Lippi, Opt. Comm. \textbf{147}, 411 (1998).
\bibitem{Chi2} For intense pulses, various processes can lead to second order susceptibilities contributions in atomic vapors as, for instance, generation of a dc electric field from multiphoton ionization of atoms resulting in a symmetry breaking. See D. S. Bethune, Phys. Rev. A \textbf{23} 3139 (1981).
\bibitem{Liao78} P. F. Liao, D. M. Bloom, and N. P. Economou, Appl. Phys. Lett. \textbf{12}, 813 (1978).
\bibitem{Harris71} J. F. Young, G. C. Bjorklund, A. H. Kung, R. B. Miles, and S. E. Harris, Phys. Rev. Lett \textbf{27}, 1551 (1971).
\bibitem{Chiao64} R. Y. Chiao, E. Garmire and C. H. Townes, Phys. Rev. Lett. \textbf{13}, 479 (1964).
\bibitem{Grischkowsky70} D. Grischkowsky, Phys. Rev. Lett. \textbf{24}, 866 (1970).
\bibitem{Lett09} R.C. Pooser, A. M. Marino, V. Boyer, K. M. Jones, and P. D. Lett, Optics Express \textbf{17}, 16722 (2009).
\bibitem{Xiao08} Y.-F. Xiao, S. K. \"O, V. Gaddam, C.-H. Dong, N. Imoto, and L. Yang, Optics Express \textbf{16}, 21462 (2008).
\bibitem{ANGELLS} F. Queiroga,W. Soares Martins, V. Mestre, I. Vidal, T. Passerat de Silans, M. Ori\'a, and M. Chevrollier, Appl. Phys. B: Lasers and Optics \textbf{107}, 313 (2012).
\bibitem{ANGELLS2} W. Soares Martins, H. L. D. de S. Cavalcante, T. Passerat de Silans, M. Ori\'a, and M. Chevrollier, Appl. Opt. \textbf{51}, 5080 (2012).
\bibitem{Sheik} M. Sheik-Bahae, A. A. Said, T.-H. Wei, D. J. Hagan, and E. W. van Stryland, IEEE J. Quantum Electronics \textbf{26}, 760 (1990).
\bibitem{Sinha02} S. Sinha, G.K. Bhowmick, S. Kundu, S. Sasikumar, S. K. S. Nair, T. B. Pal, A. K. Ray, and K. Dasgupta, \textbf{203}, 427--434 (2002).
\bibitem{Mccormick04} C.F. McCormick, D. R. Solli, R. Y. Chiao, and J. M. Hickmann, Phys. Rev. A, \textbf{69}, 023804 (2004).
\bibitem{Mccormick03} C. F. McCormick, D. R. Solli, R. Y. Chiao, and J. M. Hickmann, J. Opt. Soc. Am. B \textbf{20}, 2480 (2003).
\bibitem{Boyd} R. W. Boyd, \textit{Nonlinear Optics}, 3rd ed. (Academic, 2008).
\bibitem{Wang04} Y. Wang and M. Saffman, Opt. Comm. \textbf{241}, 513 (2004).
\bibitem{Asadpour12} S. H. Asadpour, M. Sahrai, A. Soltani, and H. R. Hamedi,  Phys. Lett. A \textbf{376}, 147 (2012).
\bibitem{Arnaud71} J. A. Arnaud, W. M. Hubbard, G. D. Mandeville, B. de la Clavière, E. A. Franke, and J. M. Franke, Appl. Opt. \textbf{10}, 2775 (1971).
\bibitem{Khosrofian83}J. M. Khosrofian and B. A. Garetz, Appl. Opt. \textbf{22}, 3406 (1983). 
\bibitem{density} The vapor particle density is obtained from the fit of a linear absorption spectrum, detected for a low intensity beam and an open-aperture configuration.
\bibitem{Steck} D. A. Steck, Cesium D line Data, http://steck.us/alkalidata/cesiumnumbers.pdf.
\bibitem{Siddons08} P. Siddons, C. S. Adams, C. Ge and I. G. Hughes, J. Phys. B: At. Mol. Opt. Phys. \textbf{41}, 155004 (2008).
 

\end{thebibliography}
\end{document}